**A Data-driven Resilience Framework of Directionality Configuration based on Topological Credentials in Road Networks**

**H M Imran Kays**
Ph.D. Candidate
School of Civil Engineering & Environmental Science
University of Oklahoma
202 W. Boyd St., Norman, OK 73019-1024
Email: kays@ou.edu

**Khondhaker Al Momin**
Ph.D. Student
School of Civil Engineering & Environmental Science
University of Oklahoma
202 W. Boyd St., Norman, OK 73019-1024
Email: momin@ou.edu

**K.K. "Muralee" Muraleetharan, Ph.D., P.E., G.E.**
Kimmell-Bernard Chair in Engineering, David Ross Boyd and Presidential Professor
School of Civil Engineering and Environmental Science
University of Oklahoma
202 W. Boyd St., Norman, OK 73019-1024
E-mail: muralee@ou.edu

**Arif Mohaimin Sadri, Ph.D.**
Assistant Professor
School of Civil Engineering & Environmental Science
University of Oklahoma
202 W. Boyd St., Norman, OK 73019-1024
E-mail: sadri@ou.edu
(Corresponding Author)

Word Count: 6,225 words + 3 table (250 words per table) = 6,975 words

*Submitted [08/01/2023]*

## ABSTRACT

Roadway reconfiguration is a crucial aspect of transportation planning, aiming to enhance traffic flow, reduce congestion, and improve overall road network performance with existing infrastructure and resources. This paper presents a novel roadway reconfiguration technique by integrating optimization based 'Brute Force' search approach and decision support framework to rank various roadway configurations for better performance. The proposed framework incorporates a multi-criteria decision analysis (MCDA) approach, combining input from generated scenarios during the optimization process. By utilizing data from optimization, the model identifies total betweenness centrality (TBC), system travel time (STT), and total link traffic flow (TLTF) as the most influential decision variables. The developed framework leverages graph theory to model the transportation network topology and apply network science metrics as well as stochastic user equilibrium traffic assignment to assess the impact of each roadway configuration on the overall network performance. To rank the roadway configurations, the framework employs machine learning algorithms, such as ridge regression, to determine the optimal weights for each criterion (i.e., TBC, STT, TLTF). Moreover, the network-based analysis ensures that the selected configurations not only optimize individual roadway segments but also enhance system-level efficiency, which is particularly helpful as the increasing frequency and intensity of natural disasters and other disruptive events underscore the critical need for resilient transportation networks. By integrating multi-criteria decision analysis, machine learning, and network science metrics, the proposed framework would enable transportation planners to make informed and data-driven decisions, leading to more sustainable, efficient, and resilient roadway configurations.

## INTRODUCTION

The increasing frequency and severity of natural disasters, extreme weather events, and unexpected disruptions create unforeseen challenges to transportation networks (*1*). The ability of transportation systems to withstand and recover from such disturbances, also referred to as network resilience, greatly concerns the urban planners, policymakers, and transportation authorities (*2*). In this context, roadway reconfiguration (*3–5*) and implementing directionality interventions (*6*) (i.e., one-way vs. two-way configuration) to improve overall network capacity have emerged as promising strategies to enhance transportation network efficiency and resilience. Here, roadway reconfiguration implies modifying the layout and connectivity of road networks to improve traffic throughput and accessibility during normal and adverse conditions. On the other hand, directionality interventions focus on managing traffic flow by adjusting lane directions (e.g., implementing reversible lanes (*7*), contra flow (*4*), optimizing traffic signal timing (*8*), etc.) to improve congestion and enhance network performance.

Under regular conditions, transportation network design is concerned with adding optimal road facilities (road segments or intersections) and determining capacity enhancement of the existing infrastructures that ensure the efficient movement of people and goods hence improving the overall resiliency (*9*). However, transportation planners face difficulties in finding appropriate policies and optimal solutions that maximize the utilization of existing resources. For example, new traffic schemes including roadway reconfiguration, signal optimization, lane balancing, along with others may be implemented in the existing traffic network that will improve the system's total travel time and externalities due to congestion without large investments in capacity improvements (*10*). Moreover, disturbance in the network created by natural disasters adds another layer of uncertainty that may destabilize the system. The consequences of which are: excessive delay, accidents, and gridlock. Such external shocks are important considerations for the resilience aspect of the network which requires planning for how much redundancy a system needs (*11*). While resiliency improvement of the system (i.e., capacity) can be constrained by improper planning due to unexpected natural hazards, roadway reconfiguration and directionality intervention can be implemented to improve the mobility, accessibility, and efficiency of the network during extreme conditions. One of the big advantages of such intervention is immediate traffic throughput and mobility improvement utilizing the existing roadway facilities. In terms of rapidity and robustness, transportation networks would get immediate benefit from this.

systematic topological and ranking network combined network adverse impacts due to This paper investigates the potential benefits of systematic roadway reconfiguration by leveraging advanced methods of directionality interventions on transportation network resilience. While previous studies searched optimal roadway configuration using traffic simulation models (*12*) and optimization algorithms (*9*), topological credentials of the network are not incorporated in the framework. The "topological credentials" of a transportation system is defined as system level quantification and ranking of network elements (e.g., road links, intersection, hubs, etc.) using network metrics (e.g., degree, centrality) that identifies the most critical and vulnerable components (e.g., bridges) (*13*). Moreover, the correlation between these credentials and the fundamental traffic variable (i.e., speed, flow, density, etc.) (*14, 15*) are not considered in previous studies which limits the understanding between optimization techniques and dynamic traffic behavior. By using optimization techniques and traffic simulation models combined with network based topological credentials, this research seeks to identify the most effective strategies to mitigate adverse impacts due to disruptions and maintain the functionality, hence improve resiliency of transportation systems during natural hazards. The specific objectives of the study include:

- Identify the relationship between network functional variables (e.g., system travel time, link flow) and topological credentials (e.g., betweenness centrality).
- Investigate how topological credentials affect the optimal road reconfiguration for maximum system performance, especially when disaster induced node/edge failure occurs.
- Developing a decision support model using multi criteria decision analysis and data driven approach.

The paper is organized as follows: Section 2 provides a comprehensive review of related literature on optimal roadway reconfiguration, directionality interventions, and network resilience. Section 3 outlines the methodology and results, including problem formulation, optimization, and multi criteria decision analysis (MSDA). A case study is included in section 4 that covers the application of the decision variables developed in MCDA along with the intervention considering the topological credential of the road network under wildfire related evacuation scenarios. Lastly, Section 5 concludes the paper by summarizing the key findings, insights, and highlighting the significance of this investigation for future efforts in enhancing transportation network resilience.

## BACKGROUND AND RELATED WORKS

Roadway reconfiguration, also known as road network reconfiguration or road layout optimization, refers to the process of modifying the design and configuration of roadways to improve traffic flow, reduce congestion, enhance safety, and optimize overall transportation network performance. Many studies explored whether it is possible to find universal solution of one-way and two-way optimal roadway configuration, but so far, there is no agreement on the global optimal organization that improve urban mobility (*16*). Some studies suggest the change from a bidirectional to unidirectional route (*17–19*), while others found better solution when the configuration change from one-way to two-way (*20–24*). However, mixed operation, like, some links are two-way and some are one-way, profoundly affects traffic congestion and improves traffic flow, leading to urban sustainability (*25*). Other advantages of partial one-way operation includes simple and coordinated signal system, higher operating speed, reduction in pedestrian crashes, etc. (*26*). Some studies considered the transportation network design configuration as Supply Design Problem where the general objective function to minimize is typically the user's total travel time (*10, 27*). Heuristic multi-criteria technique based on genetic algorithms, meta-heuristic algorithms namely Hill Climbing, Simulated Annealing, Tabu Search, Genetic Algorithms and Path Relinking are some of the optimization techniques proposed in literature to find optimal network orientation (*28*). However, these techniques are sensitive to cases and parameters governing the scenarios and finding global solutions for generalization is not possible. Roadway reconfiguration involves the process of modifying the design and configuration of roadways to improve traffic flow, reduce congestion, enhance safety, hence improving overall network performance. Many studies explored whether it is possible to find universal solution of one-way and two-way optimal roadway configuration, but so far, there is no agreement on the global optimal organization of roadway that improve urban mobility (*16*). Some studies suggest the change from a bidirectional to unidirectional traffic operation in road network (*17–19*), while others found better solution when the network configuration change from one-way to two-way (*20–24*). However, mixed operation, like, some links are two-way and some are one-way, profoundly affects traffic congestion and improves traffic flow, leading to urban sustainability (*25*). Moreover, mixed operation promote lane balancing that reduce congestion and ensure proper utilization of existing infrastructure (*29*). Other advantages of partial one-way operation includes simple and coordinated signal system, higher operating speed, reduction in pedestrian crashes, etc. (*26*). Some studies considered the transportation network design configuration as Supply Design Problem where the general objective function focuses on minimizing the user's total travel time (*10*, *27*). Heuristic multi-criteria technique based on genetic algorithms, meta-heuristic algorithms namely Hill Climbing, Simulated Annealing, Tabu Search, and Path Relinking are some of the optimization techniques proposed in literature to find optimal network orientation (*28*). However, these techniques are sensitive to cases and parameters governing the scenarios, finding global solutions for generalization is not possible.

On the other hand, network science analyzes the transportation network topology utilizing metrics from graph theory which is particularly helpful for investigating roadway configuration, especially when combined with mobility. Traditionally transportation networks are explained with network science as the properties are relevant to express through graph characteristics. Many subsystems of transportation networks like public transit, subway, multimodal infrastructure, etc. are already investigated using graph theories (*30*–*33*). The general considerations for this research are undirected and unweighted networks as the analysis does not require investigating the inward and outward data of the nodes. This method yields great simplicity where directed and weighted analysis makes it more complex (*34*). But it is more logical to consider directed networks to investigate the topological properties as in the real-world undirected networks are not always a common scenario. Moreover, the integration of accessibility, mobility, traffic flow, and network metrics are crucial for better understanding the transportation network. Guze reviewed the existing network science metrics and tried to test their applicability in combination with transportation flow theory for transportation infrastructure's safety, reliability and risk analysis (*35*). This study tested several roadway configurations; however, the findings do not suggest any solution for better roadway configuration. Likaj et al. used graph theory to find the optimal path that minimize cost using Minimum Spanning Tree graph search algorithm (*36*). However, it is crucial to identify and integrate demand with network metrics because the general consideration is that, when the demand is too high, the network tends to experience higher congestion which may be ignored for isolated network analysis. Saberi et al. investigated the urban travel demand pattern using statistical properties of weighted and directed graph networks where they developed a quantitative framework to explain the mobility of urban areas and found that travel demand networks pose similar behavior irrespective of topography and urban structure (*37*). Though topological credentials of road networks are well explored area of research, limited studies suggest the practicality of utilizing these metrics for directionality intervention for optimal roadway reconfiguration.

Because of these difficulties directed network analysis is a less explored area in the literature. The next challenge arises from traffic assignment problem which is a well-studied area in literature (*38*, *39*), however there is a literature gap on how the traffic assignment strategies can help to find better roadway configuration. Shafiei et al. have developed a dynamic traffic assignment model for large scale congested networks that can represent traffic patterns, adaptive driving behavior, travel time, and delay in the network more precisely than before (*40*). Accurate and efficient traffic assignment techniques, like, stochastic, deterministic, or dynamic traffic assignment can be utilized to capture dynamic nature of transportation system hence the link flow can be estimated from it. Such link flow is useful when analyzed together with link capacity to predict the probability of congestion which is very important to evaluate network performance. The network is likely to experience congestion if the demand reaches capacity or is not managed properly. To resolve these issues, Mahmassani et al. explored the properties of traffic flow in a congested condition and identified that the network average flow can be estimated as a nonlinear function of network average density. They also provide a deep understanding of dynamic hysteresis, the gridlock phenomenon where demand management and adaptive driving tend to reduce the size of gridlock (*41*). Saberi et al. presented a framework that explained the dynamics of congestion propagation and dissipation in a large-scale network using a simple contagion process that acts like a disease spread in the population (*42*). They also verified this interesting idea by multi-city analysis. Zhan et al. developed a framework for modeling functional failures and recoveries on complex networks which can successfully explain traffic congestion and recovery patterns on an urban scale (*43*).Aforementioned traffic assignment techniques are useful tools to explore criticality of congestion, however, these approaches do not solve or give any guideline about how to incorporate these tools to implement directionality intervention for better network performance.

Until now, it is unknown how roadway reconfiguration and change in directionality affects transportation network mobility and efficiency since the topology as well as traffic assignment alters for any change in network configuration. Exploring the relationship between functional parameters (e.g., speed, flow, density, travel time, etc.) and topological credentials (e.g., degree, centrality, etc.) of a transportation system will help to find new insights on link directionality effect on road network. Also, there remains a literature gap on how to rank different scenarios of roadway configurations and develop a decision support system which will guide the traffic managers to make informed decisions from alternative options.

**METHODS AND RESULTS**

Roadway reconfiguration and change in link directionality first affect the topological properties of the network. Subsequently, sets of shortest path for demand and traffic network flow, i.e., traffic assignment altered, hence the overall system performance changes. To find the most optimal roadway configuration, i.e., combination of one-way and two-way roads that minimize the system travel time, this study searches for all possible combinations for a small network (inset of Figure 2). Instead of defining a bi-level optimization problem, this study use 'Brute Force' search algorithm to find solutions for all possible combinations. During the search process, stochastic traffic assignment is used to estimate user equilibrium system travel time, system travel time (STT), and flow. Also, for each combination, the topological credentials are recorded for late user. After testing a total of 5,006 possible combinations, the most efficient configuration is identified (Table 1 Combination 1). However, each of 5,006 combinations are considered as an individual scenario, hence these scenarios are ranked based on STT. This data is used to train and test with machine learning, hence the weights of decision variables (i.e., total betweenness centrality, system travel time, total link traffic flow) are estimated in MCDA step. This weighted ranking equation (Equation 4) can be used by the traffic managers to rank configurations for different situations (e.g., morning and evening rush hour traffic, evacuation, etc.). Overall, the snapshot of methods used in this study is illustrated in Figure 1.

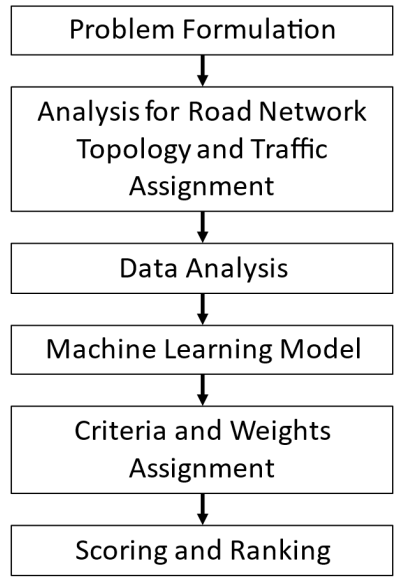

Figure 1 Flow chart showing the steps for ranking roadway configuration.

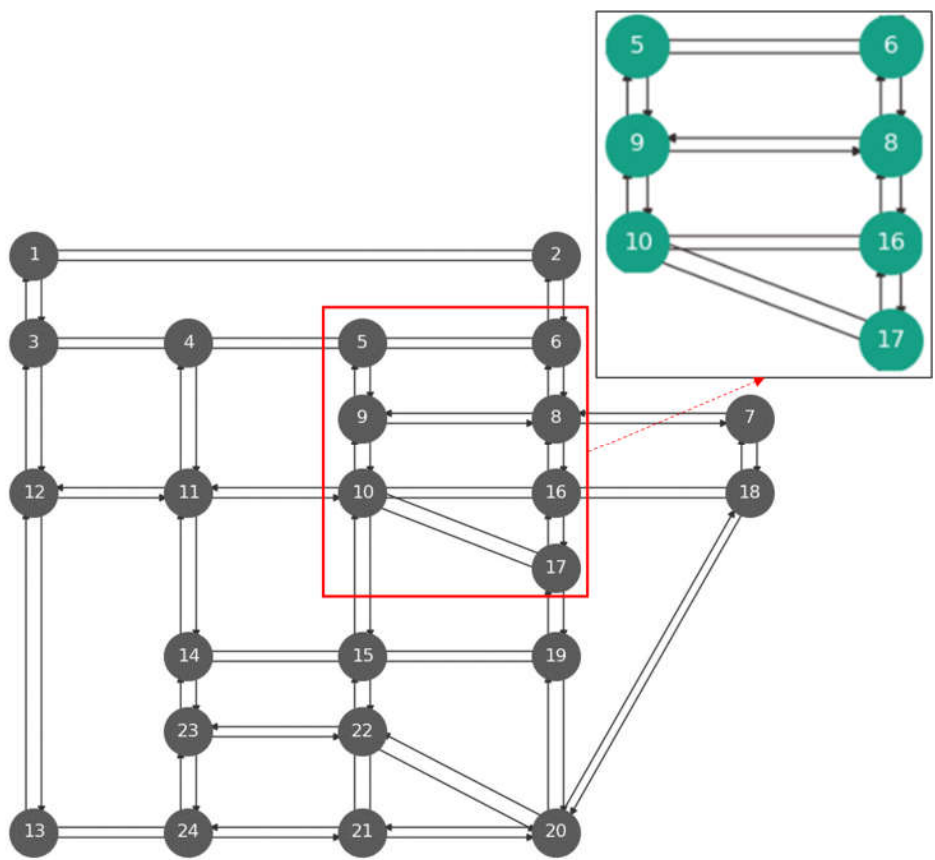

*Figure 2 Sioux Falls network and subnetwork for analysis*

## Analysis for Optimal Roadway Configuration

### Problem Formulation

The effect of directionality in a transportation network is examined in this study by utilizing a small part of the Sioux Falls network (inset network in Figure 2) which consists of 7 nodes (i.e., traffic analysis zone) and links (i.e., road segments) connecting these nodes (*44*).

Each of the links has two possible operations: one-way and two-way. To denote the flow between any two nodes, the variable '$X$' was used, while '$Y_{i,j}$' represents the direction of the road. The road orientation could be either one-way from node '$i$' to node '$j$' (denoted as 1), one-way from '$j$' to '$i$' (denoted as -1), or two-way (denoted as 0). Thus, there are three possible options for each road. Therefore, a network with '$n$' links, the total number of possible orientations can be $3^n$. To illustrate this, Figure 3(a) depicts a configuration where all roads were two-way, while Figure 3(b) shows an alternative arrangement where 4→3 is one-way, and the remaining roads are bidirectional. Consequently, the total number of possible combinations for Figure 3(b) is 34.

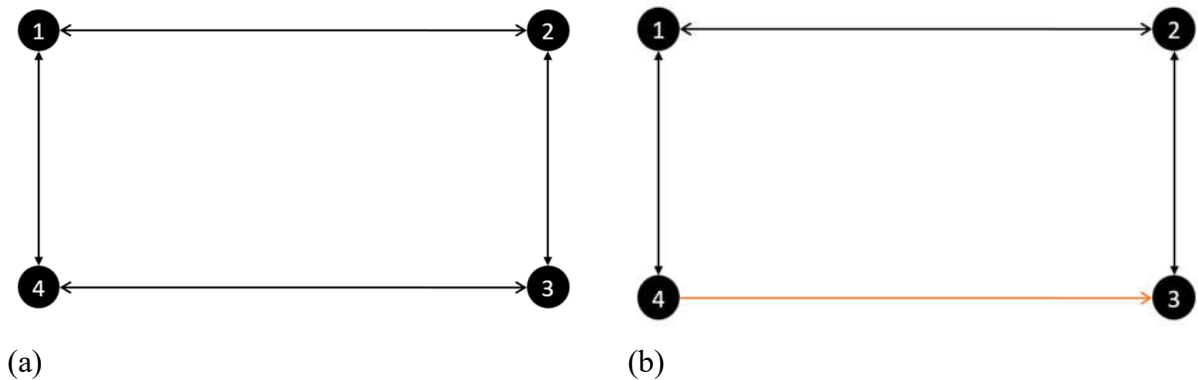

(a)                                                        (b)

*Figure 3 Concept of roadway reconfiguration*

However, when considering larger networks such as the Sioux Falls network with 38 links, the total number of possible combinations escalates to an immense value of $3^{38}$, which poses an enormous computational challenge. To evaluate the performance of each possible combination, it is necessary to assign traffic and assess the system's travel time. Assuming a computation rate of approximately 500 combinations per second, evaluating all $3^{38}$ combinations would require several hundred centuries to complete. This study aims to find a combination of road orientations in the transportation network that yields an optimal system travel time (i.e., user optimal) below a certain threshold. Here, the certain threshold is the current system travel time.

In the realm of computational complexity, a problem can be classified as an NP problem (nondeterministic polynomial-time problem) when it meets two fundamental conditions. First, it must possess verifiability, wherein a solution to the problem can be verified in polynomial time. In other words, given a proposed solution, it is possible to determine whether it is correct or not in a reasonable amount of time. Second, the problem should demonstrate the ability to be solved within nondeterministic polynomial time. Nondeterministic here means that there exists a hypothetical machine that can find the correct solution in polynomial time (*45–48*).

In this case, finding a combination of road orientations that reduces system travel time is a NP problem. It satisfies both conditions. A particular road orientation can be verified whether it reduces system travel time or not, and if there exists a hypothetical machine that can evaluate all possible combinations, then it would find an optimal road orientation that would reduce the system travel time. Therefore, this study aims to evaluate all possible combinations of a computationally feasible network, with the objective of identifying discernible patterns through the lens of network science and fundamental descriptor of traffic flow, such as edge betweenness centrality, link flow, and system travel time.

### Search for Optimal Road Configuration with Stochastic Traffic Assignment

"Brute Force" search approach (*49*) is used in this study to find the optimal solution in the road network (inset network of Figure 2) that involves systematically evaluating all possible combinations within a defined search space and identify the solution that minimizes a given objective function. The brute force method is conceptually simple and finds a guaranteed solution but can be computationally expensive and inefficient. However, being a simple network, a total of $3^9$ minus the impossible combinations is tested to find the optimal roadway configuration that ensures the STT minimization.

The concept of impossible combination is generated from the process of implementing stochastic traffic assignment for user equilibrium in this network. Stochastic traffic assignment considers uncertain or stochastic conditions like variability in travel times, demand, and other factors that affect traffic flow (*50*). This study considers, for a specific time, the demand for each node is fixed (*44*) and the traffic assignment must ensure the equilibrium that the demand is met within the analysis time. Also, this study assumes that the links that change orientation from two-way to one-way will have the full capacity (full capacity when it was two-way). However, travelers are assumed to make route choices based on shortest paths, perceived travel times and costs, which is uncertain. Different travelers in the same origin-destination (O-D) pair may choose different routes, resulting in different traffic diversions and distribution. For the "Brute Force" search approach, this study considers, if the road configuration does not have any path between the origin to destination though there is demand between the O-D, the combination is impossible. For the given network (inset network in Figure 3), out of 19,683, 5,006 possible combinations are found and tested with stochastic traffic assignment.

This study considers unweighted betweenness centrality (*BC*) as the representative of topological credential of the transportation network (*51*). BC measures the extent to which a link lies on the shortest paths between pairs of other nodes/links in the network, hence representing the importance of a link in a network. The BC of the directed network (*52*) can be expressed as equation 2.

$$C_B = \sum_{i \neq j \neq v} \frac{c_{iv}(j)}{c_{iv}} \tag{1}$$

where, $C_{iv}$ is the total number of shortest paths between node $i$ and $v$, $C_{iv}(j)$ is the number of those paths that pass through node $j$. Among others, this study particularly choose unweighted BC since this network metric is based on network's shortest path which is extensively used in transportation and resilience research (*13, 53, 54*). Also, functional metrics, like TTST, TLTF, are considered separately to evaluate network efficiency, weighting these variables in BC calculation will bias the results. Also, the topology of the network changes independently for the change in roadway configuration, as such unweighted BC can also be utilized (in the form of total betweenness centrality (TBC)) as a decision variable to find optimal roadway configuration. The effect of TBC of the network will be discussed in "Multi Criteria Decision Analysis" section (Equation 2).

This study adopted the "Brute Force" search approach with some modifications (Figure 4). In the process of finding optimal solution, the work-flow chart records the network's topological credentials as well as traffic assignment to trace the change in properties for different configuration. After 5,006 iterations, some of the most optimal configurations (Rank 5006-5004) and some of the worst configurations (Rank 2-1) with their network properties are shown in Table 1. The ranking is done based on STT which was estimated as 412,272 min for base case scenario where all the links are considered two-way. The higher the rank number, the configuration yields better STT. Also, TLTF and TBC values are included in Table 1 for comparative illustration.

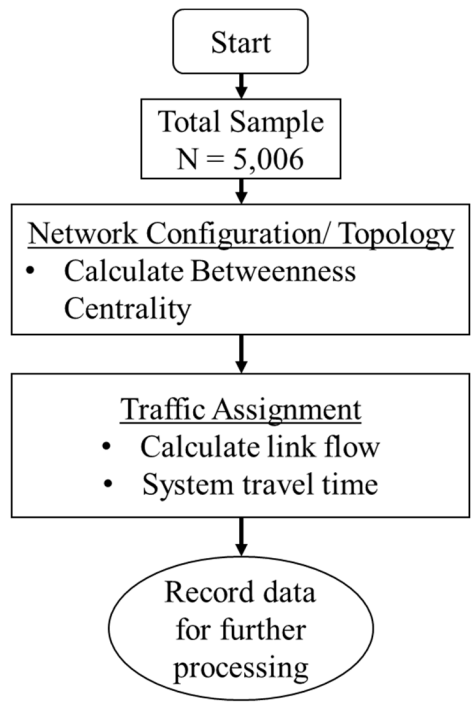

*Figure 4 The "Brute Force" search process and the optimal roadway configuration*

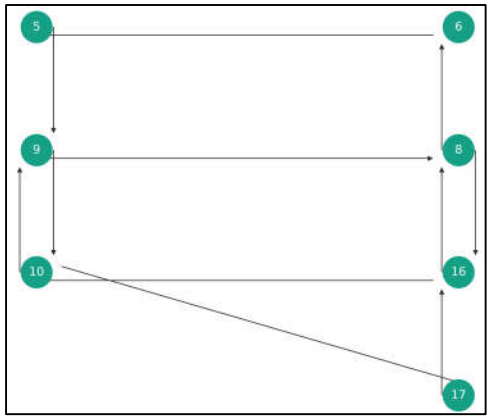

a)   Configuration 1

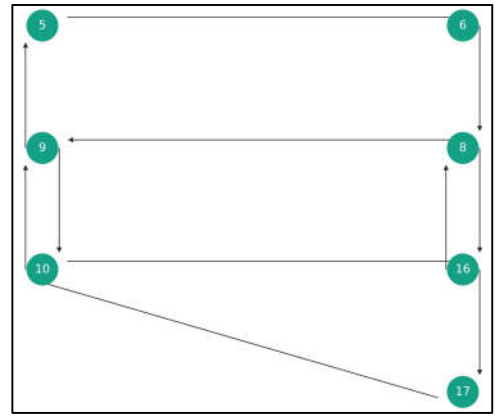

b)   Configuration 2

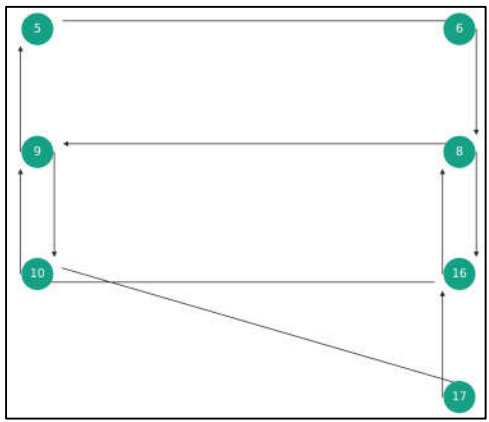

c)   Configuration 3

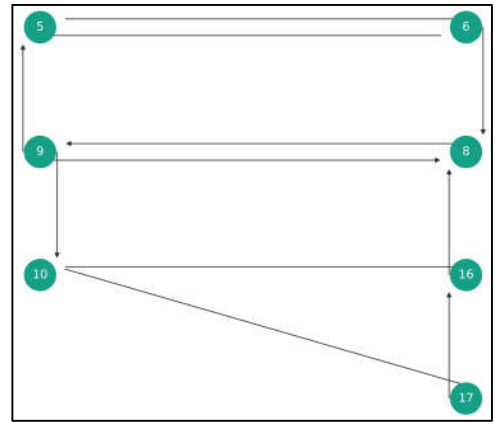

d)   Configuration 4

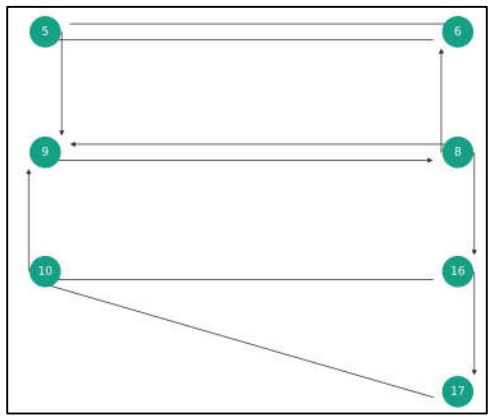

e)   Configuration 5

*Figure 5 Different roadway configuration for optimal traffic operation*

**Table 1 Comparison of five different configurations (Figure 5) including total betweenness centrality (TBC), topological credentials, total link traffic flow (TLTF), and system travel time (STT)**

| Configuration | Rank | TBC | TLTF (veh/hr) | STT (veh-hr) |
|---|---|---|---|---|
| **Configuration 1 (Figure 5a)** | 5,006 | 1.501 | 6,180,060 | 6183.41 |
| **Configuration 2 (Figure 5b)** | 5,005 | 1.503 | 6,178,680 | 6190.13 |
| **Configuration 3 (Figure 5c)** | 5,004 | 1.571 | 6,300,900 | 6239.42 |
| **Configuration 4 (Existing Network, Figure 2)** | 4,973 | 1.76 | 4,952,880 | 6871.20 |
| **Configuration 5 (Figure 5d)** | 2 | 2.714 | 8,148,840 | 1184447.00 |
| **Configuration 6 (Figure 5e)** | 1 | 2.717 | 8,095,800 | 1191498.00 |

## Multi Criteria Decision Analysis

Multi-Criteria Decision Analysis (MCDA) is a structured and systematic approach used to make complex decisions involving multiple criteria or objectives. It is a decision support tool that helps individuals or organizations to evaluate various alternatives and select the most appropriate option based on the multiple conflicting criteria (*55*). MCDA is particularly valuable when faced with decision-making situations where there is no single "best" solution, rather a range of trade-offs between competing objectives. For this study, the "Brute Force" search has already identified the optimal solution with the cost of huge computational challenge. Hence the MCDA fits the research to find solutions based on the trend in decision variables (i.e., appropriate weighting the decision variables) without much computational challenges. Moreover, these criteria will further be useful for changed network or traffic situations, e.g., disaster related evacuation.

### Criteria for Optimal Roadway Configuration

This study considers following criteria for optimal roadway configuration:

## Total Betweenness Centrality (TBC):

The total betweenness centrality (TBC) of a transportation network is defined as the sum of the BC values of all links in the network. TBC measures the overall importance or influence of all links where traffic flows based on connectivity within the network. TBC can be calculated as:

$$TBC = \sum_{n=1}^{N} BC_n \qquad (2)$$

Where, $BC_n$ is the unweighted betweenness centrality of link $n$. Another important interpretation of TBC is that this value increases as the number of links decreases in a transportation network. For example, the network in Figure 6(a) has the 10 links (highest), Figure 6(b) network has 9 links and Figure 6(c) network has 8 links. The TBC increases as the number of links reduces, for network in Figure 6(a) TBC is 1.167 with 10 links and for network in 6(c) TBC is 1.334 with 8 links. The reason is: all the shortest paths in a network follow links to reach from origin to destination, therefore a network with a smaller number of links will have a high number of shortest paths in them, hence the BC of these links increase resulting higher TBC.

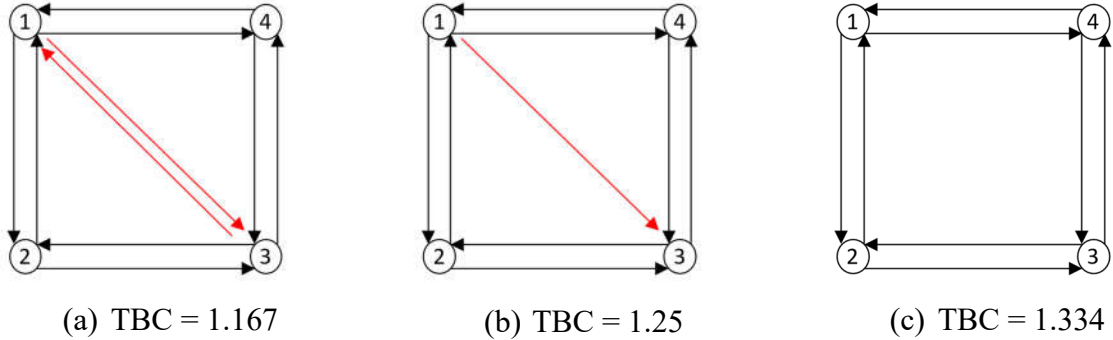

(a)  TBC = 1.167            (b)  TBC = 1.25            (c)  TBC = 1.334

*Figure 6 Change in TBC value with the change in network configuration.*

**Total Link Traffic Flow (TLTF):**

Total link traffic flow (TLTF) is the sum of the traffic flow values for all individual links which represents the overall volume of vehicles moving through the entire network, providing an important measure of the network's capacity and utilization (*56*). The total link traffic flow is expressed as:

$$TLTF = \sum_{all\ links} LTF \qquad (3)$$

Where, LTF is estimated during traffic assignment (see previous section). Total LTF can be used to identify bottleneck links and evaluate the efficiency and performance of the network under different scenarios and traffic conditions, hence this is used as a decision criterion for MCDA.

**System Travel Time (STT):**

System Travel Time in a transportation network is the summation of total time taken by each vehicle to travel from origin to a destination using shortest paths (stochastic paths for this study). It represents the overall travel experience for users traversing the transportation system and is a crucial performance metric for assessing the efficiency and effectiveness of the network. Therefore, this important traffic related metric is used as decision variable for this study.

***Determining Weights of Decision Variables using Machine Learning Technique***

This study used 5,006 observations, generated during "Brute Force" search, for training the machine learning model (ML). These observations include best to worst cases i.e., all possible roadway configuration with STT, TBC, and TLTF. These cases are ranked and assigned with a score, termed as "Directionality Configuration Score (DCS)" based on the system travel time, the lower the system travel time the higher the score. Therefor the data for machine learning model is structured as follows: 'DCS' as dependent variable, 'TBC', 'TLTF', and 'STT' as independent variables where all the variables are scaled between 0 to 1 to remove the bias of any variable in ML model. This study uses supervised learning using Ridge regression (*57*) which is a linear regression technique that adds a penalty term to the ordinary least squares cost function. Ridge regression is particularly useful when data may have multicollinearity. The data structure in this study includes TLTF and STT, both of which are estimated during stochastic traffic assignment process. Also, these variables are connected through fundamental traffic flow relations, hence there is a possibility of multicollinearity in data. Therefore, Ridge regression technique best fits this situation.

**Table 2 Manual labeling of different scenarios with DCS and data structure for ML with ridge regression**

| Directionality Configuration Score | TBC | TLTF | STT |
|:---:|:---:|:---:|:---:|
| 50 | 1.501 | 44876.97 | 371409.0 |
| 49 | 1.503 | 42295.04 | 374365.6 |
| 48 | 1.571 | 35282.30 | 383092.4 |
| 47 | 1.563 | 47780.15 | 383398.1 |
| 46 | 0.881 | 37246.71 | 389743.9 |

Moreover, Ridge regression requires the $\alpha$ parameter to be estimated, as such 8 different combinations (0.001, 0.005, 0.009, 0.01, 0.1, 0.5, 1.0, 10.0) are tested from which 0.01 generates the best results. 10-fold cross validation is used to improve the performance of the model. The model accuracy is estimated as: Mean Absolute Error= 0.092, and R-squared= 0.623, which shows the applicability of model for this study. The estimated constant term and coefficient for predicting DCS is:

$$DCS = 1.178 - 0.129 * (TBC) + 0.086 * (TLTF) - 1.180 * (\text{STT}) \qquad (4)$$

The coefficients and their signs in Equation 4 explain the effects of the decision variables on DCS. When the number of links in a network reduces (Figure 6), TBC increases and negatively affects DCS. This effect is logical in terms of resiliency of the network as reduced number of links creates less redundancy in the system, such reduction negatively affects the overall resilience. Also, total network traffic capacity is likely to be reduced, hence this coefficient comes with a negative (–) sign which indicates opposite relationship. TLTF comes with a positive (+) sign which means the network with high traffic throughput performs better, i.e., indicating high operational efficiency, therefore positively affecting the DCS. On the other hand, STT is considered as one of the influential decision variables to rank and assign DCS to each scenario (during manual labeling). Increased STT indicates that the network needs more time to move the vehicles meaning inefficiency of the system. This opposite relationship is expressed with (–) sign in Equation 4.

Equation 4 can be a great tool for decision making, particularly for comparing scenarios to make informed choices. For example, during a wildfire evacuation scenario, the topology of the transportation network changes, e.g., some links may be inaccessible, one-way operation in some link may improve the situation (contra flow). In these scenarios, the decision makers should start with reducing the TBC, one the approaches to do it can be reducing the BC of high central links by utilizing the emergency lanes or creating more routing options in the network. This way, more redundancies will be added to the system to absorb the surge of additional demand. Some of the strategies to improve redundancy can be multiple routes and modes of transport, parallel road network, emergency response plans (plan for roadway reconfiguration), multi-modal transportation integration, back-up maintenance vehicles, and so forth, which will improve the overall resiliency of the roadway network.

**Case Study: Application of DCS During Disaster**

To illustrate the application of DCS (Equation 4), this study considers hypothetical scenarios given in Figure 7, where a wildfire is approaching Node 9. An emergency warning is issued to evacuate from Node 9, hence people in Node 9 have 3 options to evacuate: 9→5, 9→8, 9→10. It is considered that before the evacuation warning, the network is configured with the most optimum orientation that ensures minimum STT (Figure 7 (a)). This optimal configuration, considered here, is estimated in the previous section (Table 1). To ensure accessibility for evacuation and avoid node isolation, the network requires minimum two-way operation in links: 5↔6, 6↔8, 8↔16 (Figure 7(b)). However, this configuration increases the overall STT for which the traffic manager needs to find another roadway configuration that improves overall traffic operation. To identify the other option, the decision variables and DCS in Equation 4 will be utilized.

At first, in Equation 4, the TLTF and STT are kept constant, the intervention is applied based on TBC. It is found that TBC increased from 1.5013 to 1.9047 after the evacuation started (Table 3). To reduce the TBC, another link 16→17 was introduced in the network and estimated TBC becomes 1.8095. To calculate the TLTF and STT, stochastic traffic assignment is used (demand remains same in nodes) and found that during evacuation (Table 3) TLTF reduced from 103,001 to 101,758 veh/min. Also, STT increases significantly from 371,005min to 1,783,663min (increase 380.77%). After the intervention (adding link 16→17) applied in the network, TLTF and STT becomes 102,588veh/min and 1,723,867min respectively. These values are normalized to calculate the DCS and found the best Configuration A (Figure 7) scored 1.127, Configuration 2 (Figure 7) scored -0.131 and Configuration 3 (Figure 7) scored 0.007. For this case study, intervention applied only based on TBC, however reconfiguration can be applied based on STT and TLTF too. The marginal effects of decision variables, discussed in previous section, are well captured by Equation 4 in this case study. When the TBC increased during evacuation, subsequent STT and TLTF increased which reduced the DCS of this scenario. After adding one directional lane 16→17, TBC reduces (from 1.094 to 1.8095), consequently STT reduces and TLTF increases which give this configuration a better score (i.e.,0.007) than evacuation scenario. From a decision maker's perspective, evacuation destabilizes the optimality of the network, but based on DCS configuration 3 can be selected for better traffic throughput in the network. In terms of resilience, the effect of redundancy can be observed in this case study, adding one lane to reduce TBC adds extra resources (i.e., capacity) in the traffic network. The system resilience will increase if other estimators remain constant.

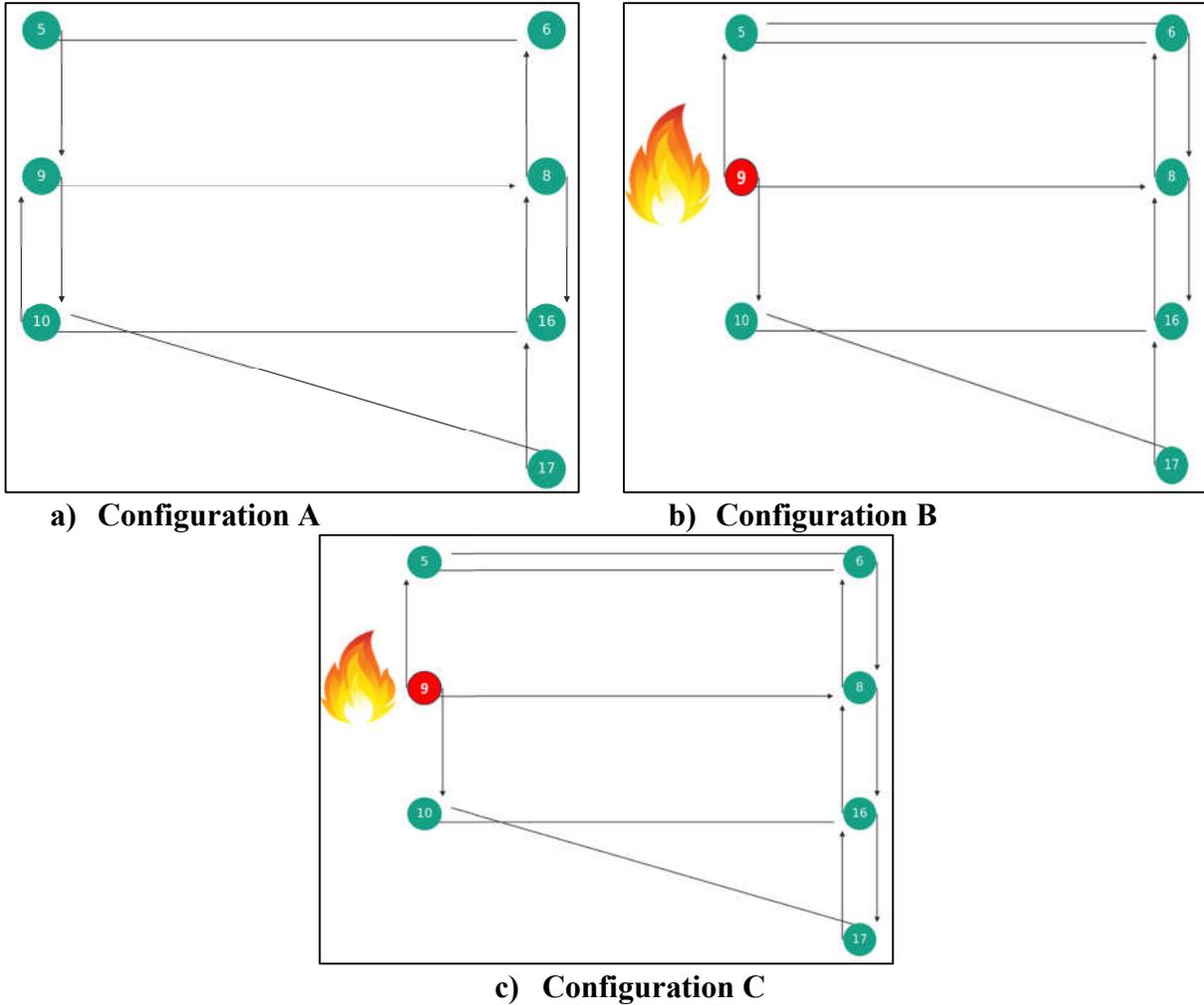

a) **Configuration A**            b) **Configuration B**

c) **Configuration C**

*Figure 7 Testing different scenarios to observe the effects of directional intervention in road network*

**Table 3 Configuration for optimal operation, Evacuation, and intervention scenarios along with the estimated decision variables and DCSs**

| Configuration | TBC | TLTF veh/hr | STT (veh-hr) | DCS | % Change in STT |
|---|---|---|---|---|---|
| **Existing Roadway Configuration (Configuration A in Figure 7)** | 1.5013 | 6180060 | 6183.42 | 1.127 | 0.00% |
| **During Evacuation (Configuration B in Figure 7)** | 1.9047 | 6105480 | 29727.72 | -0.131 | -380.77%[*] |
| **Reconfigured for Evacuation (Configuration C in Figure 7)** | 1.8095 | 6155280 | 28731.12 | 0.007 | -364.65%[*] |

*Note: 16.12% improvement from Configuration B to C after applying directionality intervention*

## CONCLUSIONS AND FUTURE SCOPE OF RESEARCH

In this study, an optimization based ('Brute Force' search) approach is used to find the most efficient roadway configuration that ensures the minimum system travel time. The advantage of optimal roadway reconfiguration is that the traffic can be managed with existing infrastructure and resources without large investments. In the optimization process, many scenarios (5,006) have been tested, which is possible due to the small size of the network. However, such optimization techniques are identified as NP problem for big transportation networks. As such, the next steps of this study look for patterns in data (5,006 observations) to develop a decision support system that ranks alternative scenarios for better traffic throughput and reduced STT. For this, multi criteria decision analysis (MCDA) concept is used to develop systematic decision support framework for rational choice. Three independent decision criteria are identified from data analysis which are crucial for ranking different scenarios. TBC is estimated from the topology of the network, which is independent but affects the overall traffic operation. Also, each new configuration changes the roadway topological credentials hence all dependent operations and network orientation changes. TLTF and STT are estimated during the process of stochastic traffic assignment. These indicators independently explain the traffic conditions and network efficiency in terms of operation. Machine learning model using Ridge regression is used to observe the effect of these indicators on the DCS for different scenarios. The ML model is trained and tested with 5,006 observations and estimated constant as well as the coefficient terms (weights) show credible results with good accuracy. The developed equations can be used as a decision support tool to choose better roadway configuration. This research generates following new insights:

- As the network size reduces (i.e., number of links reduces) TBC of links increases. The reason is that shortest paths between nodes are accessing less number links, hence the centrality value grows. As such, TBC provides a systematic understanding of directionality configurations when some links become inaccessible.
- Increased TBC of a network also relates to reduced system redundancy, i.e., the network will have less resource to be utilized during emergencies hence negatively impacts the overall resilience of the network. Therefore, a balance between TBC, TLTF, and STT will maximize the vehicle throughput in the network. This balance is achieved for the network analyzed in this study, which is not certain for other networks. This means, the developed relationship using MCDA does not guarantee scalability of the estimated weights of the decision variables.
- The most optimal configuration may not yield in the most resilient network solution as optimal configuration may lose redundancy to adopt new situations such as disruptions.

The developed decision support framework will help the agencies and traffic managers to take timely decisions about roadway configuration during normal and special occasions. For example, evacuation during hurricane or wildfire, induced demand comes, and the traffic managers can utilize this framework to adopt new configuration for better traffic management. However, the indicators used in this research are system level parameters, while more microscopic traffic parameters may have better influence on ranking equation. Moreover, sensitivity analysis is not performed due to lack of real data. Another limitation of this study comes from lack of testing the scaling effect, this research cannot answer how these predictors can assess bigger networks. Model validation for bigger networks would make this study more interesting.

Despite these challenges and limitations, the proposed decision support framework has significant takeaways for future research. Some of these include but not limited to:

- Systematic evaluation of link sequencing for directional interventions in the network that enhances the effectiveness of existing infrastructure.
- Generalization of the decision support model with the help of validation and sensitivity analysis. The validation process should consider different types of roadway configurations from different cities. Traffic simulation software can be useful for this research.
- Scalability of the decision support equation is crucial for generalization, which is not tested in this study, future studies include this scope in the research objectives.
- Reducing computational challenges by introducing more advanced and efficient optimization techniques to find better roadway configuration.

## ACKNOWLEDGMENTS

The material presented in this paper is based on work supported by the National Science Foundation under Grant No. OIA-1946093. Any opinions, findings, and conclusions or recommendations expressed in this paper are those of the authors and do not necessarily reflect the views of the National Science Foundation.

## AUTHOR CONTRIBUTIONS

The authors confirm the contributions to the paper as follows: study conception and design: H Kays, A. M. Sadri; analysis and interpretation: H Kays, K Momin, A. M. Sadri; draft manuscript preparation: H Kays, K Momin, K Muraleetharan, A. M. Sadri. All authors reviewed the results and approved the final version of the manuscript.

## REFERENCES

1. Litman, T. Lessons From Katrina and Rita: What Major Disasters Can Teach Transportation Planners. *Journal of Transportation Engineering*, Vol. 132, No. 1, 2006, pp. 11–18. https://doi.org/10.1061/(ASCE)0733-947X(2006)132:1(11).

2. Wan, C., Z. Yang, D. Zhang, X. Yan, and S. Fan. Resilience in Transportation Systems: A Systematic Review and Future Directions. *Transport Reviews*, Vol. 38, No. 4, 2018, pp. 479–498. https://doi.org/10.1080/01441647.2017.1383532.

3. A. Konstantinidou, M., K. L. Kepaptsoglou, and A. Stathopoulos. A Multi-Objective Network Design Model for Post-Disaster Transportation Network Management. *Promet - Traffic&Transportation*, Vol. 31, No. 1, 2019, pp. 11–23. https://doi.org/10.7307/ptt.v31i1.2743.

4. Kim, S., S. Shekhar, and M. Min. Contraflow Transportation Network Reconfiguration for Evacuation Route Planning. *IEEE Transactions on Knowledge and Data Engineering*, Vol. 20, No. 8, 2008, pp. 1115–1129. https://doi.org/10.1109/TKDE.2007.190722.

5. Zhang, X., S. Mahadevan, and K. Goebel. Network Reconfiguration for Increasing Transportation System Resilience Under Extreme Events. *Risk Analysis*, Vol. 39, No. 9, 2019, pp. 2054–2075. https://doi.org/10.1111/risa.13320.

6. Gayah, V. V., and C. F. Daganzo. Analytical Capacity Comparison of One-Way and Two-Way Signalized Street Networks. *Transportation Research Record*, Vol. 2301, No. 1, 2012, pp. 76–85. https://doi.org/10.3141/2301-09.

7. Wolshon, B., and L. Lambert. Reversible Lane Systems: Synthesis of Practice. *Journal of Transportation Engineering*, Vol. 132, No. 12, 2006, pp. 933–944. https://doi.org/10.1061/(ASCE)0733-947X(2006)132:12(933).

8. An Integrated Model for Evacuation Routing and Traffic Signal Optimization with Background Demand Uncertainty - Ren - 2013 - Journal of Advanced Transportation - Wiley Online Library. https://onlinelibrary.wiley.com/doi/full/10.1002/atr.1211. Accessed Jul. 31, 2023.

9. Farahani, R. Z., E. Miandoabchi, W. Y. Szeto, and H. Rashidi. A Review of Urban Transportation Network Design Problems. *European Journal of Operational Research*, Vol. 229, No. 2, 2013, pp. 281–302. https://doi.org/10.1016/j.ejor.2013.01.001.

10. Gallo, M., L. D'Acierno, and B. Montella. A Meta-Heuristic Approach for Solving the Urban Network Design Problem. *European Journal of Operational Research*, Vol. 201, No. 1, 2010, pp. 144–157.

11. Weilant, S., A. Strong, and B. M. Miller. Incorporating Resilience into Transportation Planning and Assessment. 2019.

12. Theodoulou, G., and B. Wolshon. Alternative Methods to Increase the Effectiveness of Freeway Contraflow Evacuation. *Transportation Research Record*, Vol. 1865, No. 1, 2004, pp. 48–56. https://doi.org/10.3141/1865-08.

13. Ahmed, M. A., A. M. Sadri, A. Mehrabi, and A. Azizinamini. Identifying Topological Credentials of Physical Infrastructure Components to Enhance Transportation Network Resilience: Case of Florida Bridges. *Journal of Transportation Engineering, Part A: Systems*, Vol. 148, No. 9, 2022, p. 04022055. https://doi.org/10.1061/JTEPBS.0000712.

14. Jayaweera, I., K. Perera, and J. Munasinghe. Centrality Measures to Identify Traffic Congestion on Road Networks: A Case Study of Sri Lanka. *IOSR Journal of Mathematics (IOSRJM)*, 2017.

15. Henry, E., L. Bonnetain, A. Furno, N.-E. E. Faouzi, and E. Zimeo. Spatio-Temporal Correlations of Betweenness Centrality and Traffic Metrics. Presented at the 2019 6th International Conference on Models and Technologies for Intelligent Transportation Systems (MT-ITS), 2019.

16. Bindzar, P., J. Saderova, M. Sofranko, P. Kacmary, J. Brodny, and M. Tutak. A Case Study: Simulation Traffic Model as a Tool to Assess One-Way vs. Two-Way Traffic on Urban Roads around the City Center. *Applied Sciences*, Vol. 11, No. 11, 2021, p. 5018. https://doi.org/10.3390/app11115018.

17. Study on the Influence of One-Way Street Optimization Design on Traffic Operation System - Jun Zhang, Xinxin Zhang, Yanni Yang, Bing Zhou, 2020. https://journals.sagepub.com/doi/full/10.1177/0020294020932366. Accessed Jul. 19, 2023.

18. Gilham, J. Reviewing Potential One-Way Street Conversions in Established Neighbourhoods. Presented at the Transportation 2014: Past, Present, Future - 2014 Conference and Exhibition of the Transportation Association of Canada // Transport 2014 : Du passé vers l'avenir - 2014 Congrès et Exposition de 'Association des transports du Canada, 2014.

19. How Multi-Lane, One-Way Street Design Shapes Neighbourhood Life: Collisions, Crime and Community: Local Environment: Vol 22, No 8. https://www.tandfonline.com/doi/abs/10.1080/13549839.2017.1303666. Accessed Jul. 19, 2023.

20. Riggs, W., and J. Gilderbloom. Two-Way Street Conversion: Evidence of Increased Livability in Louisville. *Journal of Planning Education and Research*, Vol. 36, No. 1, 2016, pp. 105–118. https://doi.org/10.1177/0739456X15593147.

21. Riggs, W., and B. Appleyard. The Economic Impact of One to Two-Way Street Conversions: Advancing a Context-Sensitive Framework. *Journal of Urbanism: International Research on Placemaking and Urban Sustainability*, Vol. 11, No. 2, 2018, pp. 129–148. https://doi.org/10.1080/17549175.2017.1422535.

22. Evaluating Urban Downtown One-Way to Two-Way Street Conversion Using Multiple Resolution Simulation and Assignment Approach | Journal of Urban Planning and Development | Vol 133, No 4. https://ascelibrary.org/doi/full/10.1061/%28ASCE%290733-9488%282007%29133%3A4%2822%29. Accessed Jul. 19, 2023.

23. One-Way to Two-Way Street Conversions as a Preservation and Downtown Revitalization Tool: The Case Study of Upper King Street, Charleston, South Carolina - ProQuest. https://www.proquest.com/docview/304866988?pq-origsite=gscholar&fromopenview=true. Accessed Jul. 19, 2023.

24. Lyles, R. W., C. D. Faulkner, and A. M. Syed. CONVERSION OF STREETS FROM ONE-WAY TO TWO-WAY OPERATION. 2000.

25. Karimi, H., B. Ghadirifaraz, S. N. Shetab Bousheheri, S.-M. Hosseininasab, and N. Rafiei. Reducing Traffic Congestion and Increasing Sustainability in Special Urban Areas through One-Way Traffic Reconfiguration. *Transportation*, Vol. 49, No. 1, 2022, pp. 37–60. https://doi.org/10.1007/s11116-020-10162-4.

26. One-Way/Two-Way Street Conversions. https://safety.fhwa.dot.gov/saferjourney1/Library/countermeasures/13.htm. Accessed Jul. 19, 2023.

27. Cascetta, E. Transportation Systems Engineering: Theory and Methods Kluwer Ac. *P2001*, 2001.

28. Cantarella, G. E., and A. Vitetta. The Multi-Criteria Road Network Design Problem in an Urban Area. *Transportation*, Vol. 33, 2006, pp. 567–588.

29. Xing, J., E. Muramatsu, and T. Harayama. Balance Lane Use with VMS to Mitigate Motorway Traffic Congestion. *International Journal of Intelligent Transportation Systems Research*, Vol. 12, No. 1, 2014, pp. 26–35. https://doi.org/10.1007/s13177-013-0067-7.

30. Derrible, S., and C. Kennedy. Network Analysis of World Subway Systems Using Updated Graph Theory. *Transportation Research Record*, Vol. 2112, No. 1, 2009, pp. 17–25. https://doi.org/10.3141/2112-03.

31. Derrible, S., and C. Kennedy. Applications of Graph Theory and Network Science to Transit Network Design. *Transport Reviews*, Vol. 31, No. 4, 2011, pp. 495–519. https://doi.org/10.1080/01441647.2010.543709.

32. Henry, E., A. Furno, and N.-E. El Faouzi. A Graph-Based Approach with Simulated Traffic Dynamics for the Analysis of Transportation Resilience in Smart Cities. Presented at the Transportation Research Board 98th Annual MeetingTransportation Research Board, 2019.

33. Sarker, A. A., S. Mishra, T. F. Welch, M. M. Golias, and P. M. Torrens. A Model Framework for Analyzing Public Transit Connectivity and Its Application in a Large-Scale Multi-Modal Transit Network. Presented at the Transportation Research Board 94th Annual MeetingTransportation Research Board, 2015.

34. Fagiolo, G. Directed or Undirected? A New Index to Check for Directionality of Relations in Socio-Economic Networks. *arXiv:physics/0612017*, 2007.

35. Guze, S. Graph Theory Approach to Transportation Systems Design and Optimization. *TransNav : International Journal on Marine Navigation and Safety of Sea Transportation*, Vol. Vol. 8 no. 4, 2014. https://doi.org/10.12716/1001.08.04.12.

36. Likaj, R., A. Shala, M. Mehmetaj, P. Hyseni, and X. Bajrami. Application of Graph Theory to Find Optimal Paths for the Transportation Problem. *IFAC Proceedings Volumes*, Vol. 46, No. 8, 2013, pp. 235–240. https://doi.org/10.3182/20130606-3-XK-4037.00031.

37. Saberi, M., H. S. Mahmassani, D. Brockmann, and A. Hosseini. A Complex Network Perspective for Characterizing Urban Travel Demand Patterns: Graph Theoretical Analysis of Large-Scale Origin–Destination Demand Networks. *Transportation*, Vol. 44, No. 6, 2017, pp. 1383–1402. https://doi.org/10.1007/s11116-016-9706-6.

38. Rahman, R., and S. Hasan. Data-Driven Traffic Assignment: A Novel Approach for Learning Traffic Flow Patterns Using Graph Convolutional Neural Network. *Data Science for Transportation*, Vol. 5, No. 2, 2023, p. 11. https://doi.org/10.1007/s42421-023-00073-y.

39. Cornacchia, G., M. Nanni, and L. Pappalardo. One-Shot Traffic Assignment with Forward-Looking Penalization. http://arxiv.org/abs/2306.13704. Accessed Jul. 31, 2023.

40. Shafiei, S., Z. Gu, and M. Saberi. Calibration and Validation of a Simulation-Based Dynamic Traffic Assignment Model for a Large-Scale Congested Network. *Simulation Modelling Practice and Theory*, Vol. 86, 2018, pp. 169–186. https://doi.org/10.1016/j.simpat.2018.04.006.

41. Mahmassani, H. S., M. Saberi, and A. Z. K. Urban Network Gridlock: Theory, Characteristics, and Dynamics. *Procedia - Social and Behavioral Sciences*, Vol. 80, 2013, pp. 79–98. https://doi.org/10.1016/j.sbspro.2013.05.007.

42. Saberi, M., H. Hamedmoghadam, M. Ashfaq, S. A. Hosseini, Z. Gu, S. Shafiei, D. J. Nair, V. Dixit, L. Gardner, S. T. Waller, and M. C. González. A Simple Contagion Process Describes Spreading of Traffic Jams in Urban Networks. *Nature Communications*, Vol. 11, No. 1, 2020, p. 1616. https://doi.org/10.1038/s41467-020-15353-2.

43. Zhan, X., S. V. Ukkusuri, and P. S. C. Rao. Dynamics of Functional Failures and Recovery in Complex Road Networks. *Physical Review E*, Vol. 96, No. 5, 2017, p. 052301. https://doi.org/10.1103/PhysRevE.96.052301.

44. TransportationNetworks/SiouxFalls at Master · Bstabler/TransportationNetworks. *GitHub*. https://github.com/bstabler/TransportationNetworks. Accessed Jul. 31, 2021.

45. Thomas H, C., E. Charles, R. Ronald L, and S. Clifford. Introduction to Algorithms Third Edition. Mit Press, , 2009.

46. Wooldridge, M., and P. E. Dunne. On the Computational Complexity of Qualitative Coalitional Games. *Artificial Intelligence*, Vol. 158, No. 1, 2004, pp. 27–73.

47. Cook, S. The P versus NP Problem. *Clay Mathematics Institute*, Vol. 2, 2000.

48. Garey, M. R., D. S. Johnson, and L. Stockmeyer. Some Simplified NP-Complete Problems. 1974.

49. Cevallos, F., and F. Zhao. Minimizing Transfer Times in Public Transit Network with Genetic Algorithm. *Transportation Research Record*, Vol. 1971, No. 1, 2006, pp. 74–79. https://doi.org/10.1177/0361198106197100109.

50. Daganzo, C. F., and Y. Sheffi. On Stochastic Models of Traffic Assignment. *Transportation Science*, Vol. 11, No. 3, 1977, pp. 253–274. https://doi.org/10.1287/trsc.11.3.253.

51. Assessing Robustness in Multimodal Transportation Systems: A Case Study in Lisbon | European Transport Research Review | Full Text. https://etrr.springeropen.com/articles/10.1186/s12544-022-00552-3. Accessed Jul. 31, 2023.

52. Freeman, L. C. The Gatekeeper, Pair-Dependency and Structural Centrality. *Quality and Quantity*, Vol. 14, No. 4, 1980, pp. 585–592. https://doi.org/10.1007/BF00184720.

53. Matin, A., S. Armstrong, S. Mitra, S. Pallickara, and S. L. Pallickara. Rapid Betweenness Centrality Estimates for Transportation Networks Using Capsule Networks. Presented at the 2022 Fourth International Conference on Transdisciplinary AI (TransAI), 2022.

54. Matin, A. *Towards Interactive Betweenness Centrality Estimation for Transportation Network Using Capsule Network*. M.S. Colorado State University, United States -- Colorado, 2022.

55. State-of-the-Art Review on Multi-Criteria Decision-Making in the Transport Sector - ScienceDirect. https://www.sciencedirect.com/science/article/pii/S2095756420301045. Accessed Jul. 30, 2023.

56. Hackl, J., and B. T. Adey. Estimation of Traffic Flow Changes Using Networks in Networks Approaches. *Applied Network Science*, Vol. 4, No. 1, 2019, pp. 1–26. https://doi.org/10.1007/s41109-019-0139-y.

57. McDonald, G. C. Ridge Regression. *WIREs Computational Statistics*, Vol. 1, No. 1, 2009, pp. 93–100. https://doi.org/10.1002/wics.14.